\documentclass[aps,pra,reprint,superscriptaddress]{revtex4-1}
\usepackage{graphicx, amsmath, SIunits}
\usepackage{hyperref}

\usepackage[short]{datetime}

\usepackage[usenames]{color}

\begin{document}
\title{Sub-Doppler laser cooling of $^{40}$K with Raman gray molasses on the $D_2$ line}
\date{14$^{\rm th}$ Dec 2016}
\author{G.\,D.~Bruce}
\email{gdb2@st-andrews.ac.uk}
\affiliation{University of Strathclyde, Department of Physics, SUPA, Glasgow G4 0NG, United Kingdom}
\affiliation{SUPA School of Physics and Astronomy, University of St Andrews, North Haugh, St Andrews KY16 9SS, United Kingdom}
\author{E.~Haller}
\affiliation{University of Strathclyde, Department of Physics, SUPA, Glasgow G4 0NG, United Kingdom}
\author{B.~Peaudecerf}
\affiliation{University of Strathclyde, Department of Physics, SUPA, Glasgow G4 0NG, United Kingdom}
\author{\mbox{D.\,A.~Cotta}}
\affiliation{University of Strathclyde, Department of Physics, SUPA, Glasgow G4 0NG, United Kingdom}
\author{M.~Andia}
\affiliation{University of Strathclyde, Department of Physics, SUPA, Glasgow G4 0NG, United Kingdom}
\author{S.~Wu}
\affiliation{Department of Physics, State Key Laboratory of Surface Physics and Key Laboratory of Micro and Nano Photonic Structures (Ministry of Education), Fudan University, Shanghai 200433, China}
\author{M.\,Y.\,H.~Johnson}
\affiliation{University of Strathclyde, Department of Physics, SUPA, Glasgow G4 0NG, United Kingdom}
\affiliation{SUPA School of Physics and Astronomy, University of St Andrews, North Haugh, St Andrews KY16 9SS, United Kingdom}
\author{B.\,W.~Lovett}
\affiliation{SUPA School of Physics and Astronomy, University of St Andrews, North Haugh, St Andrews KY16 9SS, United Kingdom}
\author{S.~Kuhr}
\email{stefan.kuhr@strath.ac.uk}
\affiliation{University of Strathclyde, Department of Physics, SUPA, Glasgow G4 0NG, United Kingdom}

\pacs{37.10.De,67.85.-d}

\begin{abstract}
Gray molasses is a powerful tool for sub-Doppler laser cooling of atoms to low temperatures. For alkaline atoms, this technique is commonly implemented with cooling lasers which are blue-detuned from either the $D_1$ or $D_2$ line. Here we show that efficient gray molasses can be implemented on the $D_2$ line of $^{40}$K with {\it red}-detuned lasers. We obtained temperatures of $\unit{48(2)}{\micro\kelvin}$, which enables direct loading of $9.2(3)\times 10^6$ atoms from a magneto-optical trap into an optical dipole trap. We support our findings by a one-dimensional model and three-dimensional numerical simulations of the optical Bloch equations which qualitatively reproduce the experimentally observed cooling effects.
\end{abstract}

\maketitle

\section{Introduction}
Ultracold quantum gases have attracted much interest in recent years, and have become a versatile tool to investigate strongly interacting and strongly correlated quantum systems \cite{Bloch:2008}. Cooling of an atomic gas to ultralow temperatures requires a multi-stage cooling process, which starts with a laser cooling phase in a magneto-optical trap (MOT), followed by evaporative cooling in magnetic or optical traps. Directly after collecting atoms in a MOT, lower temperatures and a higher phase space density can be achieved by sub-Doppler laser cooling techniques \cite{Dalibard:1989}. A `standard' red-detuned optical molasses allows to cool the heavier alkaline atoms Rb and Cs well below the Doppler limit.
It relies on a Sisyphus-like cooling effect in which atoms climb potential hills created by polarization or intensity gradients, thereby losing kinetic energy before being optically pumped to a lower energy level.  In the case of Li and K, the only alkalines with stable {\it fermionic} isotopes, polarization gradient cooling in a standard optical molasses is less efficient because the smaller excited state hyperfine splitting leads to a higher excitation probability of transitions other than the ones used for cooling. Nonetheless, sub-Doppler temperatures were achieved for K \cite{Modugno:1999,Taglieber:2006,Gokhroo:2011,Landini:2011} and Li \cite{Hamilton:2014} at low atom densities using a red-detuned optical molasses. In most experimental setups, the MOT uses light  near-red detuned to the $F \rightarrow F'=F+1$ cycling transition of the $D_2$ ($nS_{1/2} \rightarrow nP_{3/2}$) line ($n$ is the principal quantum number). It was also shown that Doppler cooling on the narrower linewidth $nS_{1/2} \rightarrow (n+1)P_{3/2}$ transition can achieve low temperature at high density in $^6$Li \cite{Duarte:2011,Sebastian:2014} and $^{40}$K \cite{McKay:2011}, but this technique requires lasers in the ultraviolet or blue wavelength range.

Gray molasses is another powerful method for sub-Doppler laser cooling to high densities and low temperatures \cite{Grynberg:1994,Weidemueller:1994}, and it relies on the presence of bright and dark states. A spatially varying light shift of the bright states allows moving atoms to undergo a Sisyphus-like cooling effect \cite{Dalibard:1989}, in a way that hot atoms are transferred from a dark to a bright state at a potential minimum of the bright state and back again into a dark state at a potential maximum. The coupling between the dark and bright states is velocity-selective, such that the coldest atoms are trapped in the dark states with substantially reduced interaction with the light field. Most early  experiments using gray molasses created \emph{Zeeman} dark states by using circularly polarized light on $F \rightarrow F'=F$ \cite{Hemmerich:1995,Esslinger:1996} or $F \rightarrow F'=F-1$ transitions \cite{Boiron:1995,Boiron:1996,Boiron:1998} within the $D_2$ line, and blue-detuned lasers so that the energy of bright states lay above those of the dark states. In those experiments with Rb and Cs, the large hyperfine splitting in the $P_{3/2}$ states allows a large detuning of the gray molasses laser from the other transitions (in particular from the MOT $F \rightarrow F'+1$ cycling transition) such that the gray molasses provides the dominant light-scattering process. In contrast, the poorly-resolved $P_{3/2}$ states in Li and K increase the probability of undergoing transitions on the MOT cycling transition, which was thought to limit the effectiveness of the gray molasses. Efficient implementations of gray molasses with $^{39}$K \cite{Nath:2013,Salomon:2013,Salomon:2014}, $^{40}$K \cite{RioFernandes:2012,Sievers:2015}, $^{6}$Li  \cite{Burchianti:2014,Sievers:2015} and $^{7}$Li \cite{Grier:2013} have therefore used transitions of the $D_1$ line ($nS_{1/2}\rightarrow nP_{1/2}$ transitions), requiring the use of additional lasers. In some of these experiments, cooling was found to be enhanced at a Raman resonance between the cooling and repumping light fields \cite{Nath:2013,Grier:2013,Sievers:2015}. This $\Lambda$ configuration creates additional, coherent dark states, as in velocity-selective coherent population trapping schemes \cite{Aspect:1988,Aspect:1989}.

In this work, we demonstrate that such Raman dark states can be utilized for cooling of $^{40}$K to high atom densities in a {\it red}-detuned gray molasses using only light at a frequency close to the $F \rightarrow F'=F+1$ transition on the $D_2$ line. In contrast to the $D_1$-line or narrow-line cooling schemes, our scheme requires minimal experimental overhead, because the lasers are the same as the ones used for the magneto-optical trap. Our paper is structured as follows. Firstly, we model the gray molasses by computing the energy levels and photon scattering rates of bright and dark states by numerical solution of the optical Bloch equations. We present experimental evidence that our method reaches sub-Doppler temperatures and efficiently loads atoms into an optical dipole trap. We analyze our results using a semi-classical Monte Carlo simulation and find qualitative agreement with our experimental data.

\section{1D model of red-detuned gray molasses}
\begin{figure}[ht]
\includegraphics[width=0.9\columnwidth]{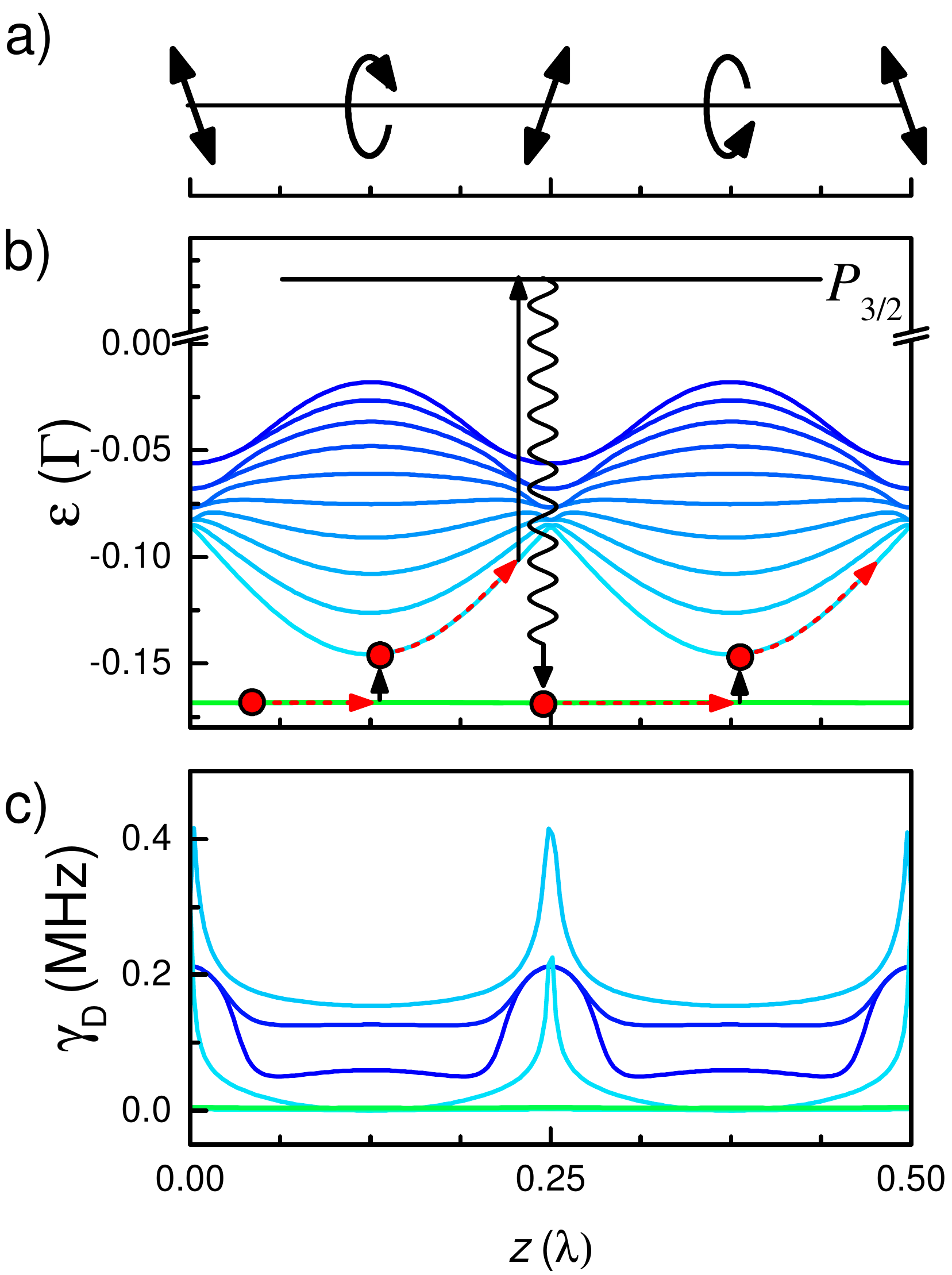}
\vspace{-0.3cm}
\caption{\label{fig:1Dmodel} Cooling mechanism in 1D $lin\perp lin$ red-detuned gray molasses ($\delta=-0.17\Gamma$). a) Variation of polarization with position $z$ over the period of the standing wave. b) Energies, $\epsilon$, of the dressed states which group in families of bright (blue) and dark states (green). An atom (red circle) in a dark state moving along $+z$ most efficiently couples to a bright state at $z=\lambda/8$. The atom undergoes Sisyphus-like cooling by travelling to an increased potential energy before an absorption followed by a spontaneous emission event returns it to a dark state at $z=\lambda/4$. c) Optical pumping rate $\gamma_{D}$, showing the dark states and the bright states corresponding to $m_{F}=-9/2, -7/2, 3/2, \text{and } 1/2$ at positions of $\sigma^{+}$ polarization (others omitted for clarity). The most likely pumping to dark states occurs at $z=\lambda/4$.}
\end{figure}

In our one-dimensional model of the gray molasses, we considered $^{40}$K atoms moving in a light field consisting of counter-propagating beams in $lin\perp lin$ configuration \cite{Dalibard:1989}. This results in constant light intensity and a polarization gradient [Fig.\,\ref{fig:1Dmodel}a)], which periodically varies from pure circular to pure linear polarization over a spatial period of half a wavelength. Each molasses beam consists of a strong cooling beam ($s=I/I_{s}=3$) and a weaker ($s=0.17$) repumper beam.
Here, $I$ is the light intensity, $I_{s} = \unit{1.75}{\milli\watt\centi\meter\rpsquared}$ is the saturation intensity of the transition and $\Gamma=\unit{2\pi\times6.035}{\mega\hertz}$ is the natural linewidth of the $D_2$ line. The cooling light is red-detuned by $\Delta=-12.3\,\Gamma$ from the $F = 9/2 \rightarrow F'=9/2$ transition and the repumper light is detuned by $\Delta - \delta$ from the $F=7/2 \rightarrow F'=9/2$ transition [see level scheme in Appendix\,\ref{app:levelScheme}, Fig.~\ref{fig:levels}b)].

Using the method outlined in Ref.\,\cite{Sievers:2015}, we calculated eigenstates and light shifts, $\epsilon$, of the dressed states (see Appendix\,\ref{app:model}), and the photon scattering rate $\gamma$ for each of the states.  The eigenstates correspond to the bare $m_{F}$ states at positions of pure $\sigma^{+/-}$ polarization [Fig.\,\ref{fig:1Dmodel}b)] and for arbitrary polarization, they are predominantly superpositions of the $m_{F}$ states within a specific $F$ manifold. The eigenstates can be grouped into dark states ($\gamma \sim 0$ at all positions) and bright states ($\gamma > 0$ at all positions). The dark states are predominantly superpositions of the $m_F$ states from the $F=7/2$ manifold, and the bright states are superpositions of $m_F$ states predominantly from the $F=9/2$ manifolds with small admixtures of states from the $F'=11/2$ manifold. The energy of the bright states is spatially varying [blue curves in Fig.~\ref{fig:1Dmodel}b)] with the dominant light-shift contribution from the $F = 9/2 \rightarrow F' = 11/2$ transition, while the light shifts of the dark eigenstates have negligible spatial variation [green curves in Fig.~\ref{fig:1Dmodel}b)]. For negative two-photon detuning ($\delta < 0$), the energy of the dark states is below the bright states [Fig.~\ref{fig:1Dmodel}b)], and above for positive detunings.

We calculated the optical pumping rate, $\gamma_{D}$, which is the rate at which an atom scatters a photon and returns to a different state (see Appendix~\ref{app:model}), at each point in space by numerical solution of the optical Bloch equations \,\cite{Sievers:2015}. Atoms are preferentially depumped from the bright states at positions of pure linear polarization  [Fig.~\ref{fig:1Dmodel}c)], which occurs at the potential maxima of the lowest-lying bright state. The probability of motional coupling  is largest when the energy difference between bright and dark states is smallest \cite{ThesisRioFernandes:2014}. By choosing $\delta$ such that the dark states have lower energy than all the bright states (in our case $\delta=-0.17\,\Gamma$), we ensure that motional coupling is strongest between the dark state and lowest-energy bright state at the potential minima.

It is critical for our scheme that the bright states belong mostly to the $F=9/2$ manifold and the dark states to the $F=7/2$ manifold, with smaller $F$. Only in this case it is possible that the energetically lowest bright state experiences the strongest light shift (at $z=\lambda/8$) while its depumping rate to the dark states is actually lowest. The main component of this bright state at $z=\lambda/8$ is from the stretched $F=9/2, m_F=9/2$ state whose scattering rate is highest, but selection rules impose that it decays mostly into the bright state manifold and not into a dark state \cite{Sievers:2015}.

\section{Experiment}
The experimental setup relevant for this study is a part of our fermionic quantum-gas microscope \cite{Haller:2015} and the cooling technique presented here is used to load atoms from a MOT into a crossed optical dipole trap (ODT). Our MOT consists of three pairs of counter-propagating circularly polarized beams containing both cooling and repumping frequencies with the same polarization.  The laser light at a wavelength of $\unit{767.7}{\nano\meter}$ for the cooling and repumper transitions is generated using two diode lasers with tapered amplifiers. Both lasers are offset-locked relative to a common master laser which is stabilized using saturated absorption spectroscopy in a vapour cell containing $^{39}$K. The six MOT beams each have $\unit{30}{\milli\metre}$ diameter and powers of up to $\unit{18}{\milli\watt}$ (cooling laser) and $\unit{0.7}{\milli\watt}$ (repumper). Cooling and repumper laser detunings are $\Delta_c = \unit{-4.6}{\Gamma}$ and $\Delta_r = \unit{-4.2}{\Gamma}$, where $\Delta_c$ is the detuning from the $F=9/2 \rightarrow F'=11/2$ transition and $\Delta_r$ is the detuning from the $F=7/2 \rightarrow F'=9/2$ transition (see level scheme in Appendix\,\ref{app:levelScheme}, Fig.\,\ref{fig:levels}a). We loaded $^{40}$K emanating from a 2D-MOT into the MOT for 4\,s and collected up to $1.3\times 10^{8}$ atoms. The atom cloud was then compressed for $\unit{10}{\milli\second}$ by changing the detunings to $\Delta_c = \unit{-1.8}{\Gamma}$ and $\Delta_r = \unit{-10.6}{\Gamma}$. All lasers and the magnetic-field gradient were  then switched off and we waited for $\unit{1.5}{\milli\second}$ to allow for eddy currents in the system to decay.

\begin{figure}[!ht]
\centering
	\includegraphics[width=0.9\columnwidth]{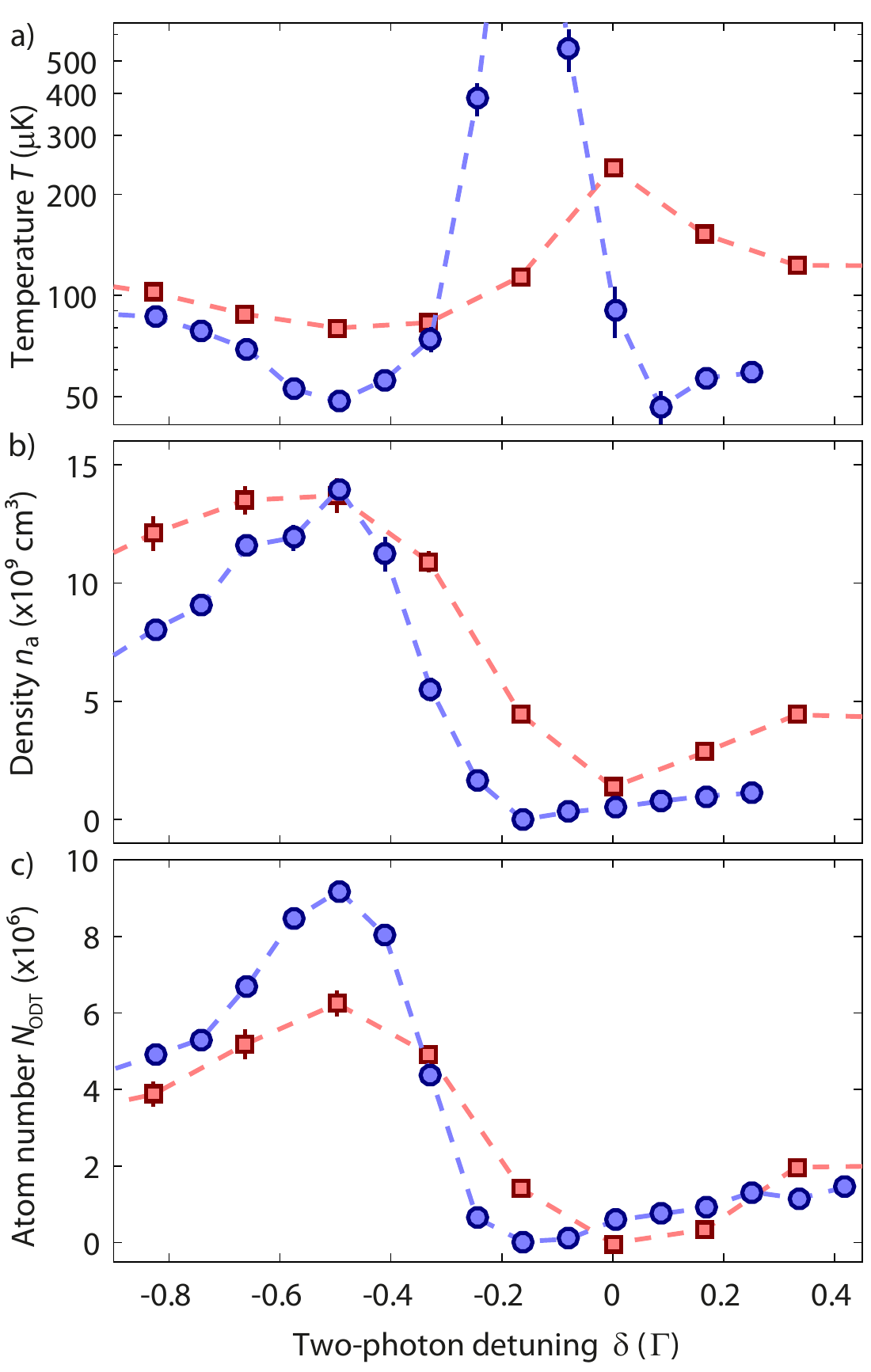}
	\caption{\label{fig:Exp1}Experimental demonstration of $D_2$-line gray molasses cooling. Dependence of a) temperature, $T$, b) free space atomic density, $n_a$, and c) number of atoms, $N_{\rm ODT}$, loaded into the optical dipole trap as a function of two-photon detuning, $\delta$, for fixed detuning $\Delta$, (red squares) and for the optimized loading sequence using a time-varying $\Delta$ (blue circles). The error bars are fit errors, and the dotted lines are to guide the eye.}
\end{figure}

For the subsequent gray molasses phase, cooling and repumper light were switched back on with powers of  $\unit{17.7}{\milli\watt}$ and $\unit{0.1}{\milli\watt}$, respectively, for a duration of 9\,ms.
The detunings of cooling and repumper beams were set such that they are now close to the Raman resonance with two-photon detuning $\delta$ and Raman detuning $\Delta \simeq -13\,\Gamma$.
We measured the temperature, $T$, of the atoms by time-of-flight absorption imaging for different values of $\delta$ [red squares in Fig.\,\ref{fig:Exp1}a)] and found a Fano-like profile with a minimum of $T=\unit{80(1)}{\micro\kelvin}$ at $\delta=-0.5\,\Gamma$ (where
the number in parentheses denotes the 1$\sigma$ uncertainty of the last
digits). The maximum density of the free atom cloud, $n_a$, after gray molasses cooling was $n_a=\unit{1.4(1) \times10^{10}}{\centi\metre\rpcubed}$ at the same detuning that yielded the lowest temperature [red datapoints in Fig.\,\ref{fig:Exp1}b)]. At this optimal two-photon detuning, atom losses during the molasses phase are negligible, whereas for higher temperatures a significant number of atoms was lost from the molasses.

%
%  Loading into ODT
%
We then loaded the atoms into an ODT, which is created by two $\unit{100}{\watt}$ laser beams from a multimode fibre laser at $\unit{1070}{\nano\meter}$ wavelength. The Gaussian beams have orthogonal linear polarization, are focused to a $\unit{300}{\micro\meter}$ 1/$e^2$ beam waist and intersect at an angle of $\unit{17}{\degree}$ at the centre of the MOT, creating a trap of  $\unit{180}{\micro\kelvin}$ depth. The ODT is switched on at full power just after the compression phase. Using the experimental procedure as described above, we loaded up to $N_{\rm ODT} = 6.2(3)\times10^6$ atoms into the ODT [Fig.\,\ref{fig:Exp1}c)], and this maximum value was found at the two-photon detuning which also yielded the lowest temperatures and highest densities in free space.

\begin{figure}[!t]
	\includegraphics[width=0.9\columnwidth]{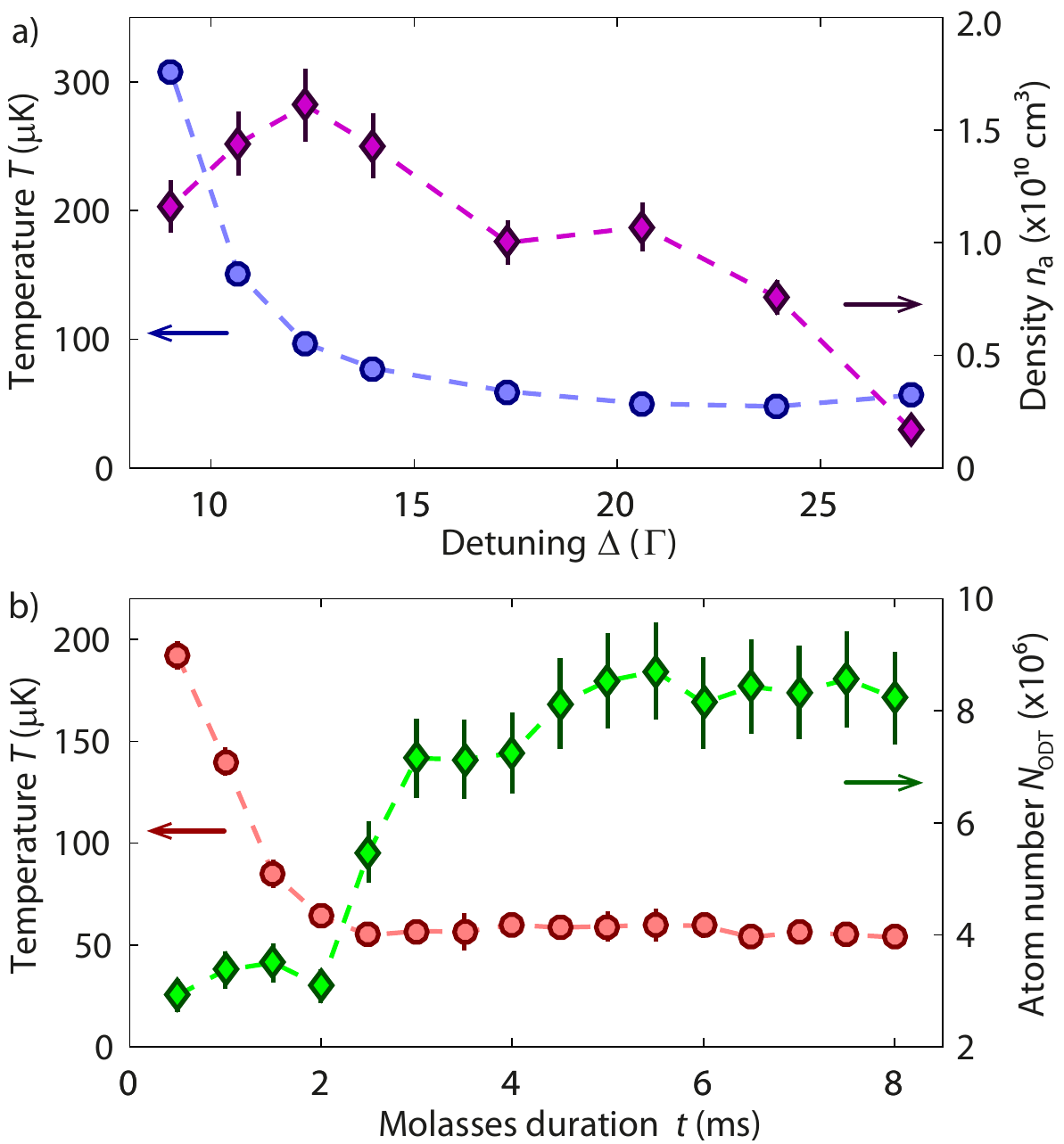}
	\caption{\label{fig:Exp2} Optimization of gray-molasses cooling. a) Temperature $T$ (blue circles) and free space atomic density, $n_a$ (purple diamonds), as a function of the one-photon Raman detuning $\Delta$ after 6\,ms of gray molasses cooling. b) Temperature $T$ (red circles) and atom number loaded into the ODT $N_{\rm ODT}$ (green diamonds) a function of the molasses duration.}
\end{figure}

In a second step, we investigated how the density changed as a function of the detuning from the excited state, $\Delta$ [Fig.\,\ref{fig:Exp2}a)]. For each value of $\Delta$, we optimized the temperature and density by varying $\delta$. We found that by moving farther away from resonance, both the temperature and density are reduced. While the reduction in temperature is desirable, the reduction in density reduces the loading efficiency into the ODT. In order to achieve both lower temperatures and higher densities, we ramped both $\Delta$ and $\delta$ during the molasses phase. We found that optimal  loading was achieved by ramping $\Delta$ from $-10.8\Gamma$ to $-20.3\Gamma$ and $\delta$ from $-0.81\Gamma$ to $-0.56\Gamma$ over a duration of 6\,ms followed by an additional hold time of 3\,ms, leading to temperatures of $\unit{48(2)}{\micro\kelvin}$ [blue circles in Fig.\,\ref{fig:Exp1}a)].

When varying the duration of the gray molasses phase [Fig.\,\ref{fig:Exp2}], we found the decrease in temperature occurred during the first 2\,ms before a steady state was reached, whereas the number of atoms in the dipole trap kept increasing and saturated after about 5\,ms of cooling. We could load up to $N_{\rm ODT}=9.2(3)\times 10^6$ atoms into the optical dipole trap, which corresponds to 7\% of the initial atom number in the molasses.  This value is limited by the volume overlap of the ODT and the molasses and it could in principle be increased by using more laser power for the ODT, which would allow us either increase the trap depth or use larger beam diameters for the same trap depth.

Our one-dimensional model can also explain the heating seen for $\delta=0$. At this detuning, the energy of the dark states is higher than the energy of the bright states, and motional coupling occurs preferentially to the most energetic bright state at the potential maxima. The atom will proceed to a lower potential before absorption and spontaneous emission pumps it to the dark state, at which point it will have \emph{gained} kinetic energy. This process is the inverse of Sisyphus cooling.

\section{3D numerical simulation}

\begin{figure}[!b]
\vspace{-0.2cm}
	\includegraphics[width=0.9\columnwidth]{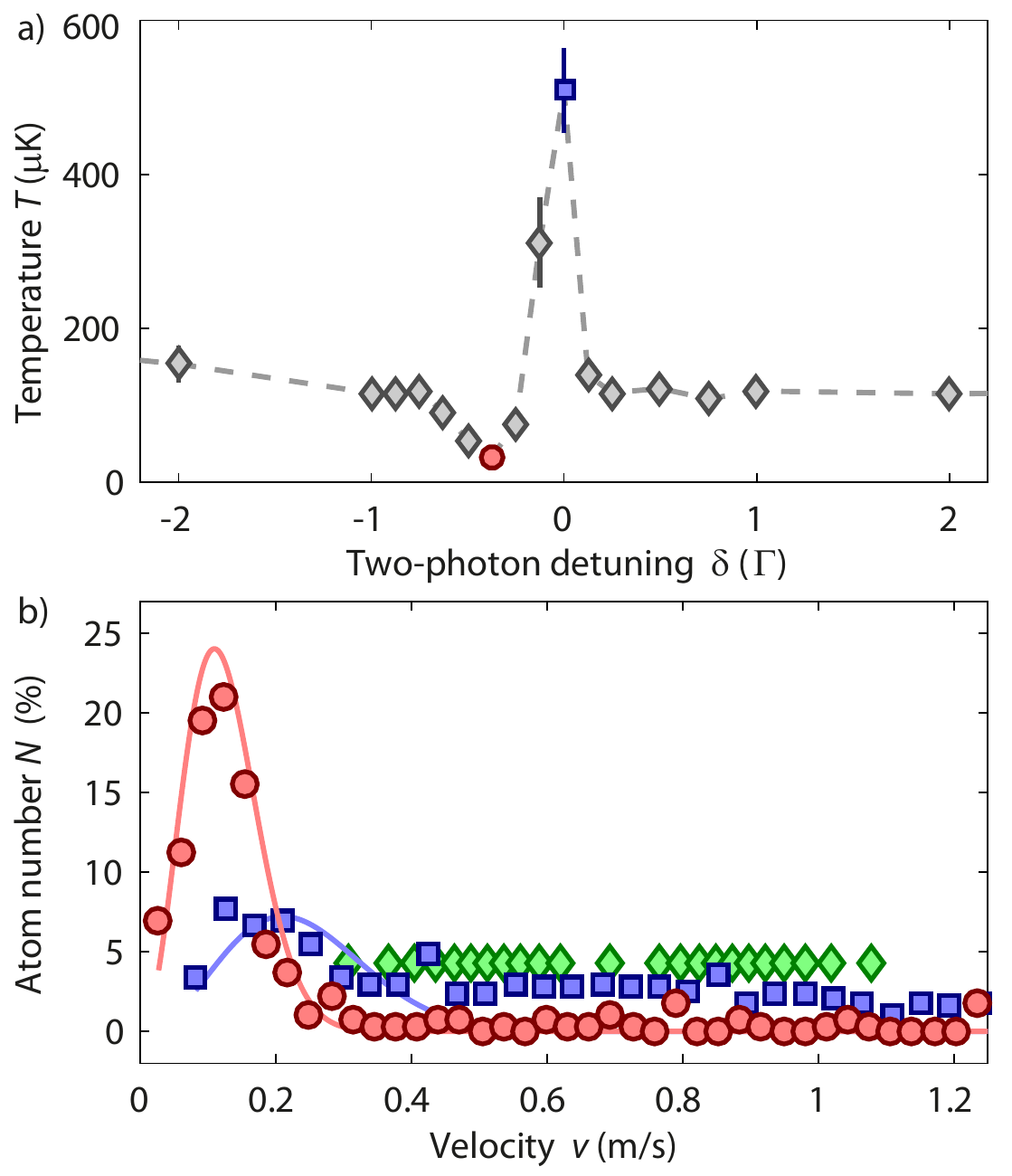}
	\caption{\label{fig:3Dmodel}
Semi-classical Monte-Carlo simulations of cooling in a 3D optical molasses. a) Dependence of temperature $T$ vs two-photon detuning $\delta$ resulting from the numerical simulations. The highlighted red and blue data points are calculated from the velocity distributions shown in b).
b) An initial non-thermal velocity distribution (green diamonds) is narrowed by Raman gray molasses cooling at $\delta = -0.4\,\Gamma$ (red circles). The red line is a fit with a Maxwell-Boltzmann distribution yielding $T = 29(1)\,\micro\kelvin$. For $\delta=0$ (blue squares), the velocity distribution broadens, and is not well described by a Maxwell-Boltzmann distribution (blue line). }
\end{figure}

To quantitatively model the cooling effect in our experimental setup, we have performed a three-dimensional numerical simulation using the same semi-classical Monte-Carlo wavefunction method detailed in Ref.~\cite{Sievers:2015}. As in the experiment, the light field in the simulation consists of three orthogonal pairs of counter-propagating laser beams in $\sigma^{+} - \sigma^{-}$ configuration, which results in both polarization and intensity gradients. Our simulation computes the coherent evolution of all of an atom's classical external states and its internal quantum states in timesteps of $\unit{1}{\micro\second}$ as it travels through the light field for a total simulation time of $\unit{10}{\milli\second}$. The internal states coherently evolve as governed by the optical Bloch equations until a randomly generated photon absorption and subsequent emission event, at which point the state is projected onto one of the ground state eigenstates. The classical external states are updated at instances of photon absorption and emission, which cause a momentum kick, but also continuously via a force proportional to the expectation value of the Hamiltonian. Each atom originates at the centre of intersection of the three pairs of laser beams. If it travels more than 15~mm ($1/e^2$ radius of our beams) from this point, in any direction, the atom is considered to have been lost.

We ran these simulations for 10\,ms duration at $\Delta = -12.3 \Gamma$ and varied the values of the two-photon detuning, $\delta$, for a fixed initial velocity of $v=0.7$\,m/s corresponding to a temperature of $T = 800\,\micro$K using the equipartition theorem $\tfrac{3}{2}k_BT = \tfrac{1}{2}m \left<v^2\right> $. The temperature extracted using the final velocity distribution and the equipartition theorem [Fig.\,\ref{fig:3Dmodel}a)] as a function of $\delta$ shows a dependence very similar to the experimental data presented in Fig.~\ref{fig:Exp1}. Far away from the two-photon resonance, the temperature of the atoms reached $180(67)\,\micro$K, close to the Doppler temperature of $145\,\micro$K.  For $\delta = -0.4\,\Gamma$ and $\delta = 0$, we simulated the time-evolution of an ensemble of atoms with an initial non-thermal distribution of velocities between 0.3\,m/s and 1.1\,m/s.  At $\delta = -0.4\,\Gamma$ the final velocity distribution is fit with a Maxwell-Boltzmann distribution (with temperature as the only free parameter) which yields $T = 29(1)\,\micro$K, which indicates cooling and thermalization [Fig.\,\ref{fig:3Dmodel}b)]. On two-photon resonance, ($\delta = 0$), poor cooling and no thermalization was observed and the corresponding temperature was $460(60)\,\micro$K. The simulation results are in accordance with both the experimental data and one-dimensional model, and we observe efficient cooling when the dark states are tuned below the bright states (now $\delta = -0.4\,\Gamma$ due to the increased light-shift from the additional beams).

\section{Summary and Conclusions}
In summary we have presented a method to realize efficient gray molasses cooling using red-detuned lasers on the $D_2$ line of fermionic $^{40}K$. The achieved temperatures and densities allow for direct loading of atoms into an optical dipole trap. A key advantage of our technique that is uses the same lasers as for the magneto-optical trap and it does not require additional lasers near the $D_1$ line.

\section*{Acknowledgments}
  We acknowledge support by EU (ERC-StG FERMILATT, SIQS, QuProCS, Marie Curie Fellowships to E.H. and B.P.), Scottish Universities Physics Alliance (SUPA), and the Leverhulme Trust (G.B., grant no. RPG-2013-074). Numerical simulations were performed using the high-performance computer ARCHIE-WeSt (\url{www.archie-west.ac.uk}) at the University of Strathclyde, funded by EPSRC grants EP/K000586/1 and EP/K000195/1.

\bibliography{D2Molasses}

\appendix

\section{40K level scheme} \label{app:levelScheme}
The atomic level scheme of the $D_2$ line of $^{40}$K and the relevant laser fields and detunings used for our cooling scheme are shown in Fig.\,\ref{fig:levels} below. Cooling light at frequency $\omega_c$ is red-detuned by $\Delta$ from the $F=9/2 \rightarrow F'=9/2$ transition, while the repumping light ($\omega_r$) is detuned by $\Delta_r = \Delta-\delta$ from the $F=7/2 \rightarrow F'=9/2$ transition. Here $\delta$ is the detuning of the two-photon Raman transition relevant for the gray molasses scheme.

\begin{figure}[!h]
	\includegraphics[width=0.8\columnwidth]{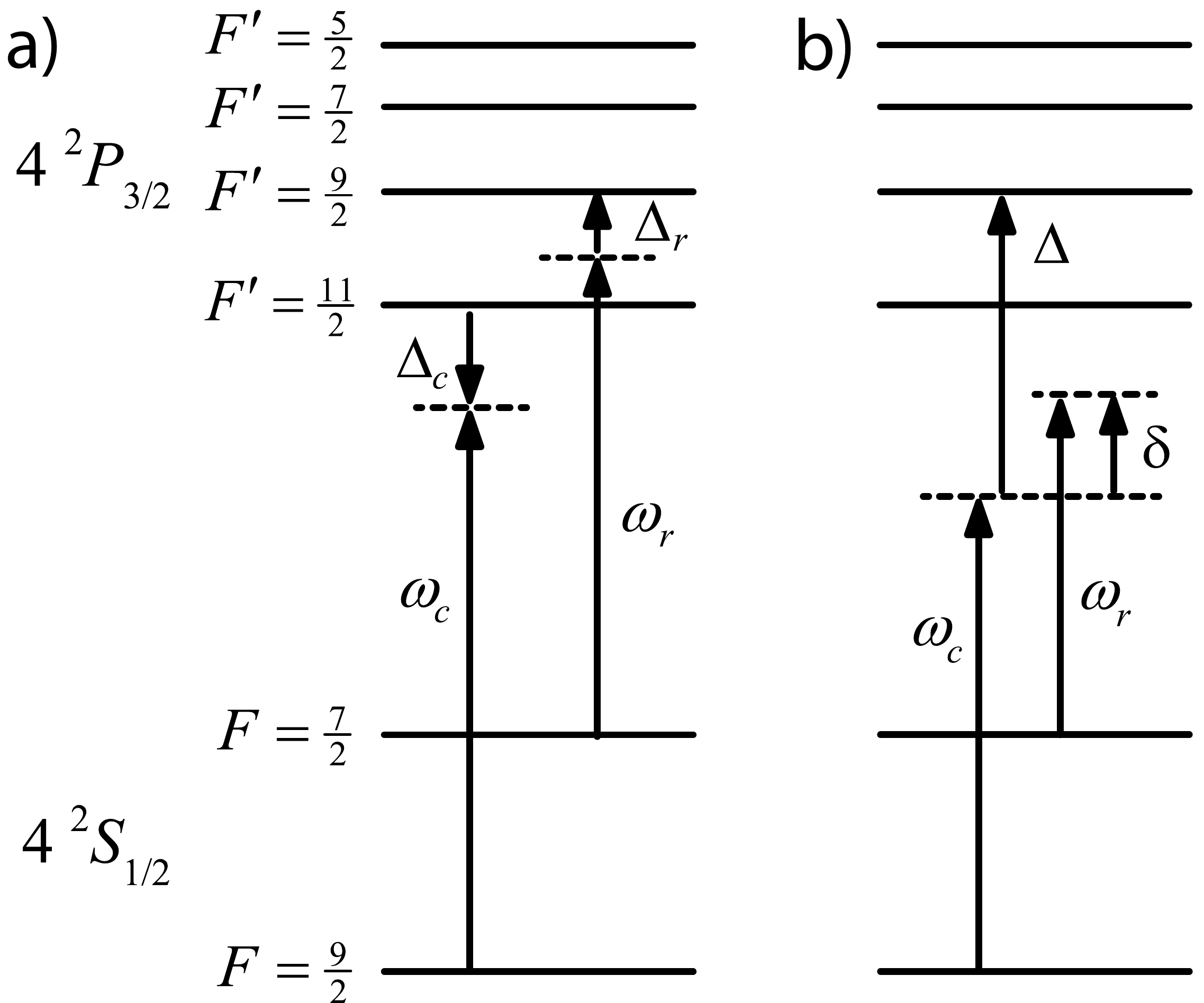}
	\caption{\label{fig:levels} Atomic level scheme of then $^{40}$K $D_2$-line together with the relevant laser fields and detunings for a) MOT operation % ($\Delta_{\rm Hfs} = 43.4$\,MHz)
and b) gray molasses cooling.}
\end{figure}

\section{1D gray molasses modelling} \label{app:model}
In our one-dimensional model of the gray molasses cooling (which follows the methods of Ref.\,\cite{Sievers:2015}), we consider a stationary atom at position $z$ within a 1D $lin\perp lin$ bichromatic optical molasses, consisting of both cooling and repumper frequencies. The cooling and repumper light field is decomposed into $\sigma^{+}$ and $\sigma^{-}$ polarizations for each beam, with quantization axis along $z$.

We find the energy eigenvalues of the dressed states (eigenstates) of the atom in the light-field by diagonalizing the Hamiltonian
\begin{equation}
 H=H_0+H_{C}.
\end{equation}

\noindent Here
\begin{eqnarray}
 \label{eqn:H0}
 \frac{H_0}{\hbar}
 &=& \sum_m \left|F=7/2,m\right> \delta \left< F=7/2,m\right| \nonumber\\
 &&+ \sum_{F' m'} \left|F',m'\right> (\delta_{{\rm hfs}, F'}-\Delta) \left< F',m'\right|,
\end{eqnarray}
is the Hamiltonian of the free atom, and $\delta_{{\rm hfs},F'}$ the hyperfine splitting of the excited state ($\delta_{{\rm hfs},F'} = 0$ for $F'=11/2$).

The light-atom coupling Hamiltonian is
\begin{eqnarray}
 H_{C}
   &=\hbar\sum_{\overset{F,m,\sigma,}{\mbox{\tiny \emph{F',m'}}}}
   & \Bigl[  c_{F,m,\sigma,F',m'} \Omega_{F,\sigma}(z)\nonumber\\
   &&\times \left|F,m\right> \left<F',m'\right|+h.c.\Bigr],
\end{eqnarray}
\noindent where $\Omega_{F,\sigma}(z)$ are the position-dependent Rabi frequencies and $c_{F,m,\sigma,F' m'}$ the relevant Clebsch–-Gordan coefficients for the transitions. The off-resonant coupling between the cooling (repumper) laser and atoms in the $F=7/2$ ($F=9/2$) ground state was neglected due to the large hyperfine splitting of the ground states.

The total scattering rate, $\gamma$, of each eigenstate at each position is the natural linewidth multiplied by the excited-state fraction of that eigenstate. The optical pumping rate $\gamma_D$ is the total scattering rate minus the rate at which the atom undergoes an absorption and emission cycle and returns to the same eigenstate (elastic scattering).

\clearpage

\end{document}